\documentclass{PoS}
\usepackage{graphics}
\usepackage{subfigure}
\usepackage{epsfig}

\input{babarsym}
\newcommand{\xai}{$\xi (2220)$}
\newcommand{\x}{$X(3872)$}
\newcommand{\y}{$Y(3940)$}

\title{\bf\boldmath{Search for the \xai\ and Study of the \x\ at \babar\/}}

\ShortTitle{The $\xi(2220)$ and the \x\/}

\author{\speaker{Arafat Gabareen Mokhtar}\thanks{On behalf of the \babar\ Collaboration}\\
        SLAC National Accelerator Laboratory\\
        E-mail: \email{mokhtar@slac.stanford.edu}}

\abstract{The \babar\ Collaboration performed a search for \xai\
production in the initial-state radiation process $e^+e^-\to \gamma
J/\psi$, $J/\psi\to \gamma K^+K^-$ or $J/\psi\to \gamma
K^0_SK^0_S$. No evidence for the \xai\ resonance has been found. The
90$\%$ confidence level upper limits on the product of branching
fractions are sensitive to the spin and helicity hypotheses. These
upper limits are of the order $10^{-5}$, below the values reported in
previous experiments. Also at \babar\/, the decays $B\to
J/\psi\pi^+\pi^-\pi^0 K$ are studied to search for the decay \x\/$\to
J/\psi\omega$. This search yields a four standard deviation evidence
for \x\/$\to J/\psi\omega$, with product branching fractions of
${\cal{B}}(B^+\to X(3872)K^+)\times {\cal{B}}(X(3872)\to
J/\psi\omega$) $=[0.6\pm0.2\stat \pm 0.1\syst ]\times 10^{-5}$, and
${\cal{B}}(B^0\to X(3872)K^0)\times {\cal{B}}(X(3872)\to
J/\psi\omega$) $=[0.6\pm0.3\stat \pm 0.1\syst ]\times 10^{-5}$. A
detailed study of the $\pi^+\pi^-\pi^0$ mass distribution from
$X(3872)$ decay favors a negative-parity assignment but does not rule
out the positive-parity hypothesis.}

\FullConference{35th International Conference of High Energy Physics - ICHEP2010,\\
		July 22-28, 2010\\
		Paris France}

\begin{document}
\section{Introduction}
The \xai\ resonance is a glue-ball candidate whose existence is not
yet established. The \x\ has been observed in several decay modes and
by several Collaborations. However, the nature of the \x\ is still not
yet understood. We present the \babar\ results on the search for
\xai\/ in radiative \jpsi\ decays~\cite{delAmoSanchez:2010hk}, and on
the evidence for the decay \x\/$\to
J/\psi\omega$~\cite{delAmoSanchez:2010jr}.

\section{Search for \xai\ in Radiative \jpsi\ Decays}
In 1986, the Mark III Collaboration
reported~\cite{Baltrusaitis:1985pu} a narrow resonance with a mass of
$\sim 2.2$ \gevcc\ in the radiative decay $J/\psi\to \gamma$\xai\/,
\xai\/$\to K^+K^-$ and \xai\/$\to K^0_SK^0_S$. A 3.6 and 4.7 standard
deviation significance for $J/\psi\to \gamma K^+ K^-$ and $J/\psi\to
\gamma K^0_S K^0_S$ modes were reported. The BES Collaboration also
reported evidence for the \xai\ in \jpsi\ radiative decays at a
comparable level of significance~\cite{Bai:1996wm}. Moreover, there
are indications for a similar structure in $\pi^-p$ and $K^-p$
collisions~\cite{Bolonkin:1987hh,Aston:1988yp,Alde:1986nx}. On the
other hand, searches for \xai\ in $p\bar{p}$
collisions~\cite{Amsler:2001fh,Evangelista:1998zg}, or two photon
production~\cite{Benslama:2002pa,Acciarri:2000ex}, have been
inconclusive.

In a recent \babar\ search~\cite{delAmoSanchez:2010hk}, the
initial-state radiation (ISR) events $e^+e^-\to \gamma_{\mathrm{ISR}}
J/\psi$, $J/\psi\to\gamma KK$ ($KK$ indicates $K^+K^-$ or
$K^0_SK^0_S$), were studied to search for the \xai\/. The \babar\ data
sample is equivalent to an integrated luminosity of 460 fb$^{-1}$,
recorded at or slightly below 10.58 \gev\/.

\begin{figure}
  \begin{center}
    \includegraphics[height=0.25\textheight,width=0.4\textwidth]{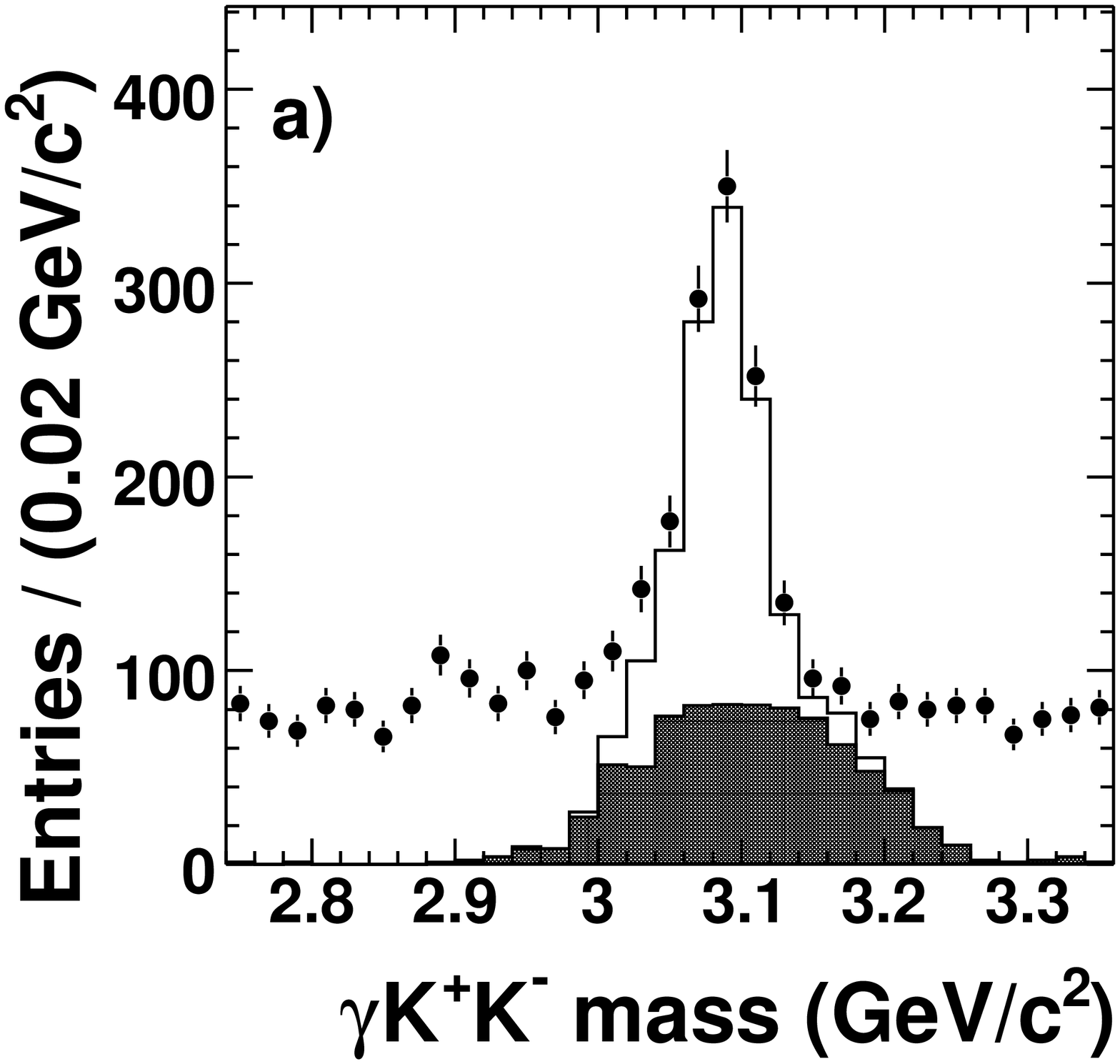}
    \includegraphics[height=0.25\textheight,width=0.4\textwidth]{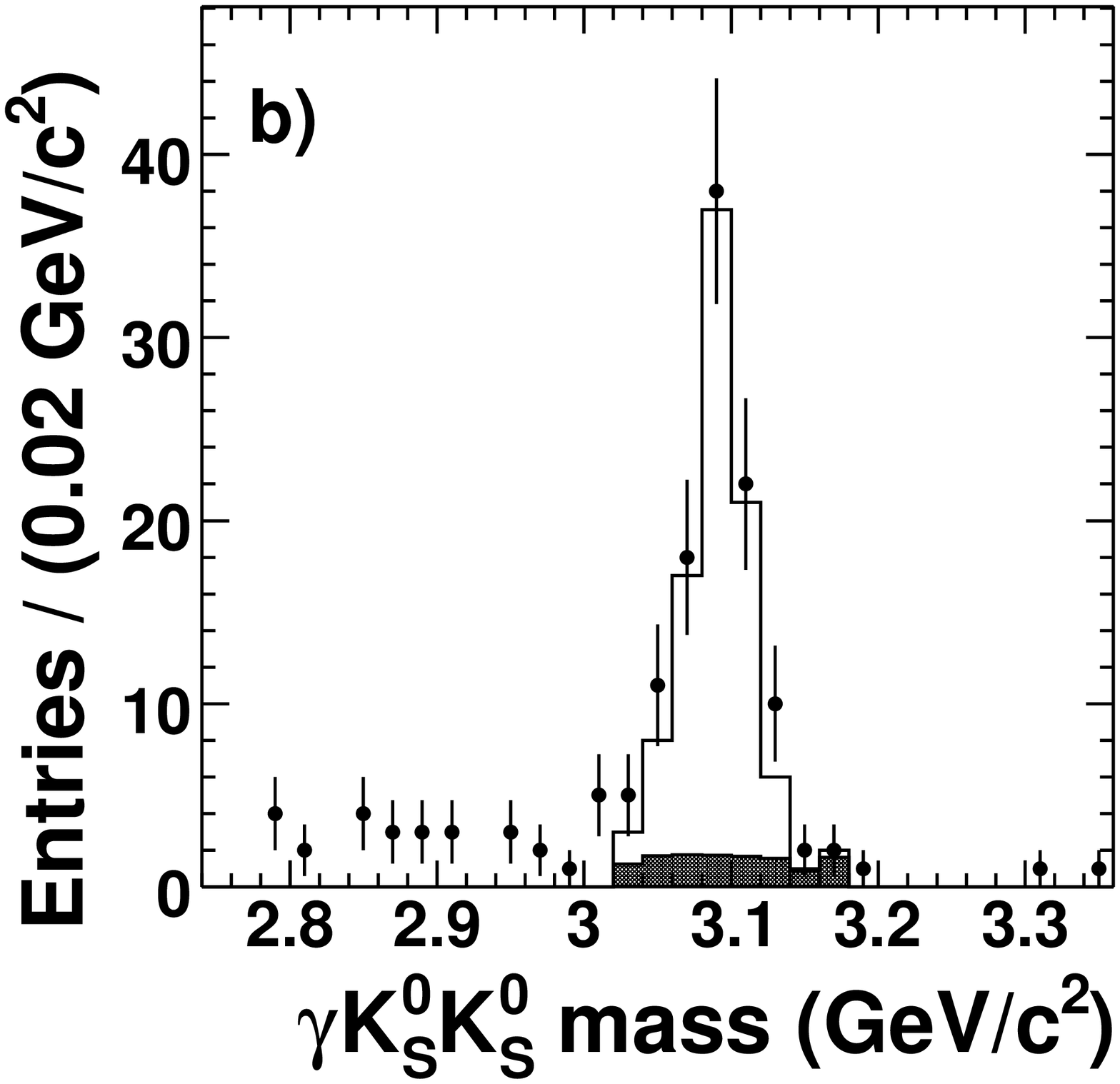}
    \caption{The mass distribution of (a) $\gamma K^+K^-$ and (b)
    $\gamma K^0_S K^0_S$ for the final sample. The dots represent the
    data and the histograms show the fits to the data when requiring a
    fit probability above 0.01. The shaded histograms represent the
    estimated background.}
    \label{fig1}
  \end{center}
\end{figure}

 The $\gamma K^+K^-$ and $\gamma K^0_S K^0_S$ mass distributions are
shown in Fig.~\ref{fig1}, where a large \jpsi\ signal is observed in
both decay modes. The background under the signal arises mainly from
partially reconstructed $J/\psi\to KK X$ or $e^+e^-\to
q\bar{q}\gamma_{\mathrm{ISR}}$ events, where $X$ can be any final
state system and $q=u,d,s,c$. The $\gamma KK$ candidates are required
to originate from a common vertex and are kinematically constrained to
the \jpsi\ nominal mass.  Each $K^0_S$ candidate in the decay
$J/\psi\to \gamma K^0_SK^0_S$ is reconstructed from two oppositely
charged tracks identified as pions. The photon emitted from the \jpsi\
has a minimum energy of 300 \mev\/.

\begin{figure}
  \begin{center}
  \includegraphics[height=0.25\textheight,width=0.4\textwidth]{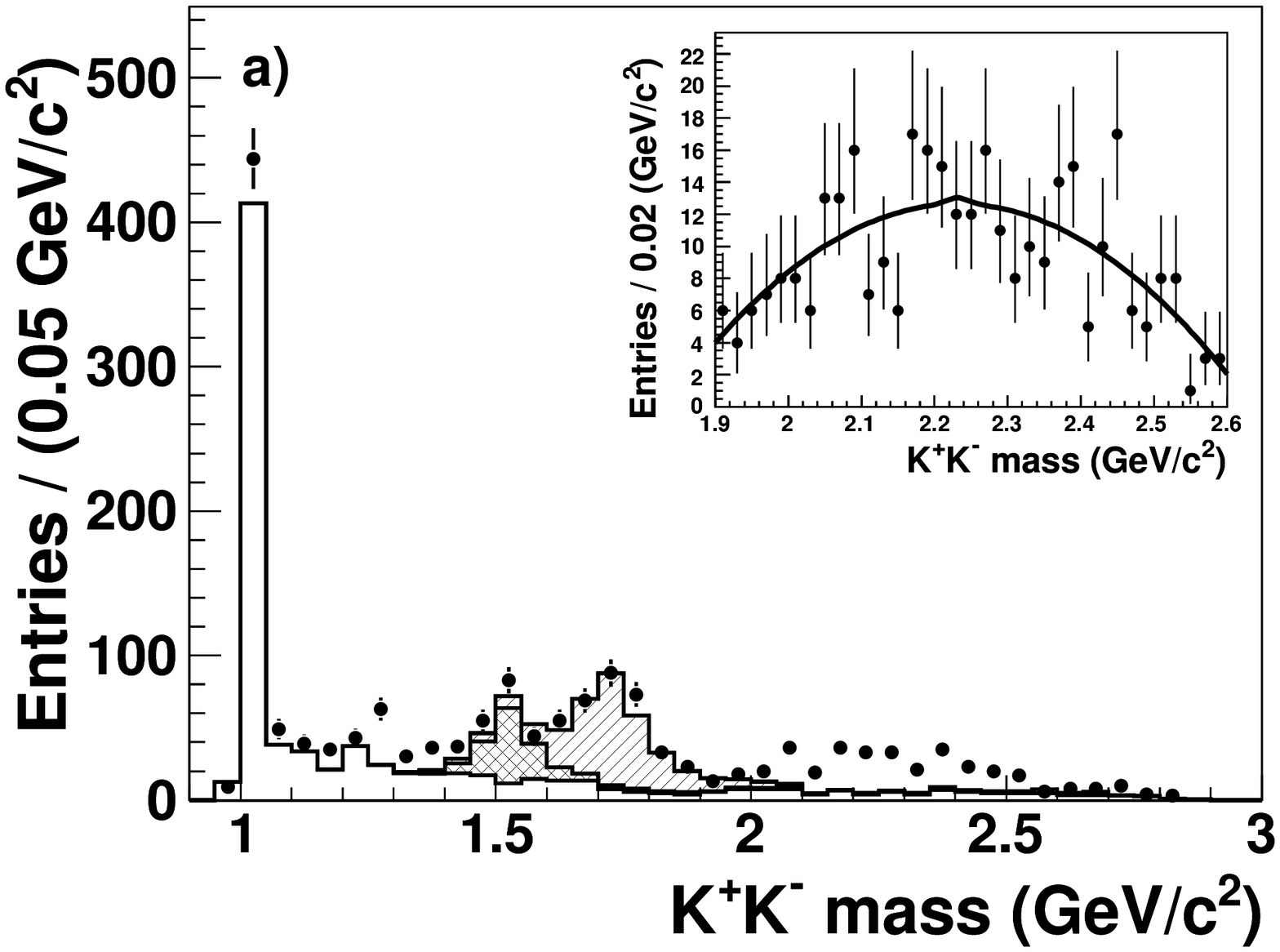}
  \includegraphics[height=0.25\textheight,width=0.4\textwidth]{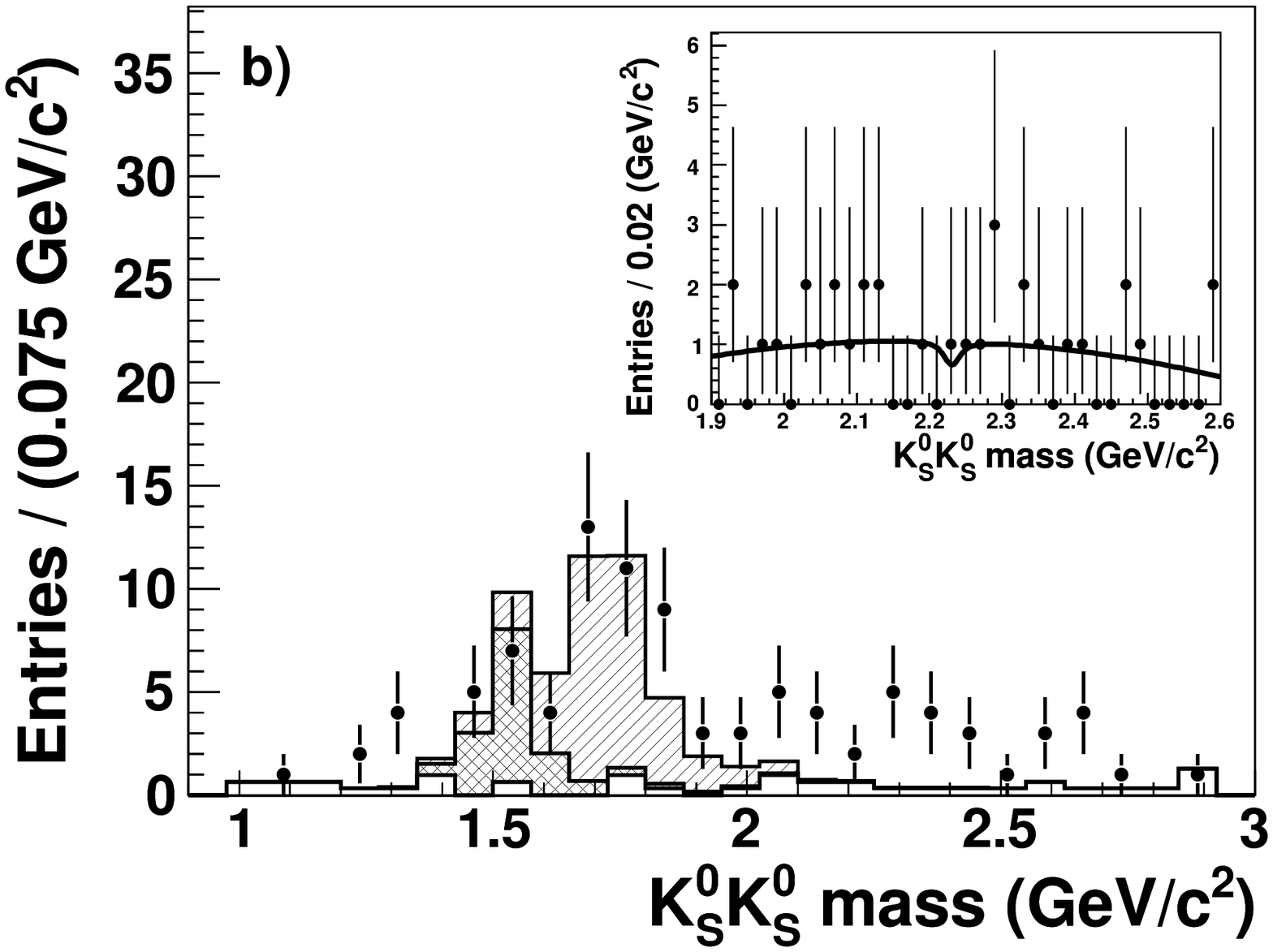}
  \caption{The fitted mass distribution for (a) $K^+K^-$ and (b)
  $K^0_SK^0_S$. The contributions of the inclusive background (open
  histograms), $J/\psi\to \gamma f_2^{\prime}(1525)$ (cross hatched
  histograms), and $J/\psi\to \gamma f_0(1710)$ (hatched histograms)
  are shown. The insets show the fit results in the \xai\ region. }
  \label{fig2}
  \end{center}
\end{figure}

The $K^+K^-$ and $K^0_SK^0_S$ mass distributions are shown in
Fig.~\ref{fig2}. The inclusive background and background events
corresponding to $J/\psi\to \gamma f_2^{\prime}(1525)$ and $J/\psi\to
\gamma f_0(1710)$, are present. The small data excess at $\sim 1.25$
\gevcc\ in the charged mode may be due to the process $J/\psi\to
\rho^0\pi^0$, with $\rho^0\to\pi^+\pi^-$, where both pions are
misidentified as kaons, and one of the photons from the $\pi^0$ is
undetected. To extract the \xai\ yield, unbinned-maximum likelihood
fits in the range $1.9\leq m_{KK}\leq 2.6$ \gevcc\ are performed. The
signal is described as a Breit-Wigner function convolved with a
Gaussian resolution function. The background is parametrized as a
second-order Chebychev polynomial. Both the mass and width of the
\xai\ are fixed. There is no evidence for \xai\ state. The upper
limits on the product of branching fractions depend on the spin and
helicity assignment. For all hypotheses of spin and helicity, the
$90\%$ confidence level upper limits for the \jpsi\/$\to
\gamma$\xai\/, \xai\/$\to KK$ product branching fractions are in the
range $(1.2-3.6)\times 10^{-5}$, smaller or close to the values
reported by the Mark III Collaboration.

\section{Evidence for \x\/$\to J/\psi\omega$}
With the discovery~\cite{Choi:2003ue} of the \x\ by the Belle
Collaboration in 2003, interest in charmonium spectroscopy has been
renewed. Confirmation of this state was obtained by CDF, D0, and
\babar\
experiments~\cite{Acosta:2003zx,Abazov:2004kp,Aubert:2004ns,Aubert:2005zh,Aubert:2008gu}. Since
then, several other charmonium-like states have been
discovered~\cite{Brambilla:2010cs}. The \x\ is the most-studied state
and the only one which has been identified in more than one decay
mode, assuming that the reported $X$, $Y$, and $Z$ states are actually
different states. A great deal of effort has been expended to
understand the nature of the \x\ especially its spin-parity assignment
($J^{PC}$). So far, $J^{PC}=1^{++}$ or $2^{-+}$ can be assigned to the
\x\/. The radiative decays \x\/$\to\gamma
J/\psi$~\cite{Abe:2005ix,Aubert:2006aj,:2008rn} and
\x\/$\to\gamma\psi(2S)$~\cite{:2008rn} indicate positive $C$
parity. At \babar\/, no charged-partner for the \x\ has been
observed~\cite{xcharged}. This establishes $I=0$.

In a previous \babar\ analysis~\cite{Aubert:2007vj} of $B\to
J/\psi\omega K$ decays, the observation of the $Y(3940)$ meson in the
decay $Y(3940)\to J/\psi\omega$, as reported by the Belle
Collaboration~\cite{Abe:2004zs}, was confirmed. In this analysis,
$\omega\to\pi^+\pi^-\pi^0$ ($\omega\to 3\pi$) candidates were required
to satisfy $0.7695\leq m_{3\pi}\leq 0.7965$ \gevcc\/, and no evidence
for the decay $X(3872)\to J/\psi\omega$ was found.

\begin{figure}
  \begin{center}
    \includegraphics[height=0.3\textheight,width=0.4\textwidth]{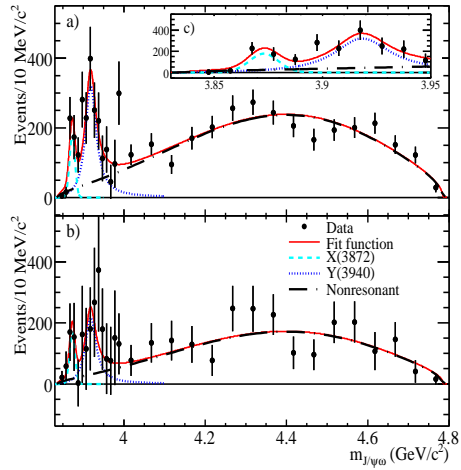}
    \caption{The $J/\psi\omega$ mass distribution for (a) $B^+\to
      J/\psi\omega K^+$ and (b) $B^0\to J/\psi\omega K^0_S$ decays;
      (c) shows the region $m_{J/\psi\omega}<3.95$ \gevcc\ of (a). The
      curves show the fit results and the individual fit
      contributions.}
    \label{fig3}
  \end{center}
\end{figure}

In a more recent \babar\ analysis~\cite{delAmoSanchez:2010jr} the same
decay mode $B\to J/\psi\omega K$ has been revisited using a slightly
larger dataset and extending the range of the $\omega$-mass region to
$0.74\leq m_{3\pi}\leq 0.7965$ \gevcc\/. All other selection criteria
are the same as in the previous analysis~\cite{Aubert:2007vj}. The
efficiency as a function of $m_{J/\psi\omega}$ varies between 5 and
7$\%$, and the mass resolution degrades from 6.5 \mevcc\ to 9
\mevcc\/, over the accessible mass range. The $J/\psi\omega$ mass
($m_{J/\psi\omega}$) distribution, after background subtraction, shows
a clear signal corresponding to $Y(3940)\to J/\psi\omega$, and
evidence for $X(3872)\to J/\psi\omega$. These signals are present in
both $B^+$ and $B^0$ samples~\cite{conjugate} as shown in
Fig.~\ref{fig3}. The $m_{J/\psi\omega}$ distributions are fitted
simultaneously after correcting for efficiency and branching
fractions. The function used in the fit has three components: an \x\
component which is a Gaussian function with fixed $\sigma=6.7$
\mevcc\/; a \y\ contribution described by a relativistic $S$-wave
Beit-Wigner function; and a nonresonant contribution given by a broad
Gaussian function multiplied by $m_{J/\psi\omega}$. The \y\ and
nonresonant components are multiplied by the phase space factor $p q$,
where $p$ is the kaon momentum in the $B$ rest frame and $q$ is the
\jpsi\ momentum in the $J/\psi 3\pi$ system. A good fit is obtained
($\chi^2/NDF=54.7/51$). The fit results are summarized in
Table~\ref{tab:results}.

\begin{table}
\begin{center}
\begin{tabular}{ll}\hline
  Quantity & Measurement \\ \hline
  Mass \x\/  (\mevcc\/) & $3873.0_{-1.6}^{+1.8}\pm 1.3$ \\
  Mass \y\/  (\mevcc\/) & $3919.1_{-3.4}^{+3.8}\pm 2.0$ \\
  Width \y\/ (\mev\/)   & $31_{-8}^{+10}\pm 5$          \\
  ${\cal{B}}(B^0\to X(3872)K^0)\times {\cal{B}}(X(3872)\to J/\psi\omega)$ ($10^{-5}$) & $0.6\pm 0.3\pm 0.1$ \\
  ${\cal{B}}(B^+\to X(3872)K^+)\times {\cal{B}}(X(3872)\to J/\psi\omega)$ ($10^{-5}$) & $0.6\pm 0.2\pm 0.1$ \\
  ${\cal{B}}(B^0\to Y(3940)K^0)\times {\cal{B}}(Y(3940)\to J/\psi\omega)$ ($10^{-5}$) & $2.1\pm 0.9\pm 0.3$ \\
  ${\cal{B}}(B^+\to Y(3940)K^+)\times {\cal{B}}(Y(3940)\to J/\psi\omega)$ ($10^{-5}$) & $3.0_{-0.6}^{+0.7}$ $_{-0.3}^{+0.5}$ \\
  ${\cal{B}}(B^0\to J/\psi\omega K^0)$ ($10^{-4}$) & $2.3\pm 0.3\pm 0.3$ \\
  ${\cal{B}}(B^+\to J/\psi\omega K^+)$ ($10^{-4}$) & $3.2\pm 0.1_{-0.3}^{+0.6}$ \\
  $R_X$ (ratio of $B^0$ to $B^+$ branching fraction to $B\to X(3872)K$) & $1.0_{-0.6}^{+0.8}$ $_{-0.2}^{+0.1}$ \\
  $R_Y$ (ratio of $B^0$ to $B^+$ branching fraction to $B\to Y(3940)K$) & $0.7_{-0.3}^{+0.4}\pm 0.1$ \\
  $R_{\mathrm {NR}}$ (ratio of $B^0$ to $B^+$ branching fraction to nonresonant $J/\psi\omega K$) & $0.7\pm 0.1 \pm 0.1$ \\
\hline
\end{tabular}
\caption{Results obtained from the most recent \babar\ analysis of
$B\to J/\psi\omega K$ decays~\protect\cite{delAmoSanchez:2010jr}.}
\label{tab:results}
\end{center}
\end{table}

When combined with the product branching fraction for $B\to X(3872)K$,
$X(3872)\to J/\psi \pi^+\pi^-$ ~\cite{Aubert:2008gu}, the \babar\
ratio of branching fractions ${\cal{B}}(X(3872)\to
J/\psi\omega)/{\cal{B}}(X(3872)\to J/\psi\pi^+\pi^-)$ has the value
$0.7\pm 0.3$ and $1.7\pm 1.3$ (combined uncertainties) for $B^+$ and
$B^0$, respectively. These results provide an average ratio of $0.8\pm
0.3$, which is in agreement with the Belle result~\cite{Abe:2005ix} of
$1.0\pm 0.4\pm 0.3$.

To judge whether the $3\pi$ originate from $\omega$ decays or not,
$3\pi$ events in the mass range of $\omega$ and $\eta$ signals are
selected. The sum of the $\omega$-Dalitz-plot
weights~\cite{Aubert:2007vj} is consistent with the number of $3\pi$
events around the $\omega$ signal. The same sum for the events around
$\eta$ signal is consistent with zero. The sum for the weighted $3\pi$
mass distribution associated with the \x\ is consistent with the
number of events observed. This justifies the $\omega$ interpretation
of the events in the \x\ region.

\begin{figure}
  \begin{center}
  \includegraphics[height=0.3\textheight,width=0.4\textwidth]{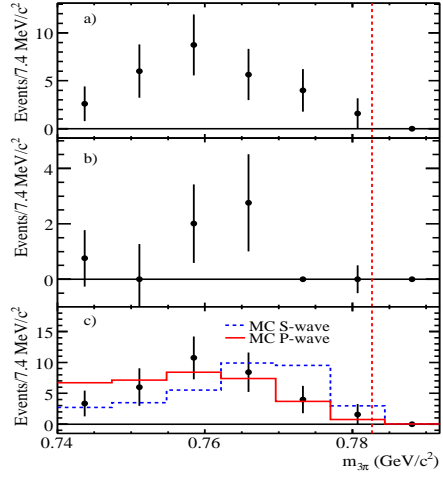}
  \caption{The $m_{3\pi}$ distribution for events that satisfy
  $3.8625\leq m_{J/\psi\omega}\leq 3.8825$ \gevcc\ for (a) $B^+$, (b)
  $B^0$, and (c) combined. The vertical line shows the $\omega$
  nominal mass. In (c), the solid (dashed) histogram shows the
  $P$-wave ($S$-wave) Monte Carlo events normalized to the number of
  data events.}
  \label{fig4}
  \end{center}
\end{figure}

The events with $3.8625\leq m_{J/\psi\omega}\leq 3.8825$ \gevcc\ are
selected for further investigation of the \x\ parity. For those
events, the $m_{3\pi}$ distributions are shown in Fig.~\ref{fig4} and
compared with the Monte Carlo simulation for different spin
assignment. The $P$-wave assignment is favored ($\chi^2/NDF=3.53/5$)
over the $S$-wave ($\chi^2/NDF=10.17/5$), hence $J^P=2^-$ is favored
over $J^P=1^+$, but the latter cannot be ruled out. Clearly this
analysis would benefit greatly from the much larger datasets available
from future facilities such as the Super $B$-factories.

\section{Acknowledgments}
I would like to thank my \babar\ Collaborators, especially Vera
L$\ddot{\mathrm{u}}$th, Bill Dunwoodie, and Bryan Fulsom, for their
contributions to the presentation at the ICHEP meeting, and to these
proceedings.


\begin{thebibliography}{99}
\bibitem{conjugate} The use of charge conjugate reactions is implied
throughout.
\end{thebibliography}
\end{document}